\documentclass[%
reprint,
amsmath,amssymb,
aps,
]{revtex4-2}

\usepackage{graphicx}
\usepackage{dcolumn}
\usepackage{bm}
\usepackage[colorlinks=true,citecolor=blue,linkcolor=blue,urlcolor=blue]{hyperref}

\usepackage{tikz}
\usepackage{xcolor}
\usepackage{soul}

\newcommand{\orcidicon}[1]{\href{https://orcid.org/#1}{\includegraphics[height=\fontcharht\font`\B]{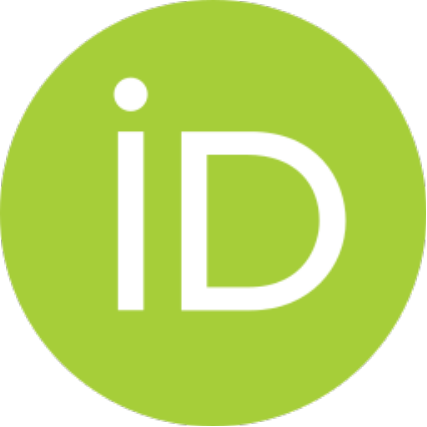}}}

\usepackage{mathrsfs}
\usepackage{pstricks}
\usepackage{color}
\usepackage{slashed}
\usepackage{epsfig}
\usepackage{amsfonts}
\def\beq{\begin{equation}}
\def\eeq{\end{equation}}
\def\bea{\begin{eqnarray}}
\def\eea{\end{eqnarray}}

\def\Re{\textrm{Re}}

\def\x{\mathbf{x}}

\def\q_perp{\mathbf{q}_{\perp}}
\begin{document}

\title{ The vacuum energy with non-ideal boundary conditions \\ via an approximate functional equation}
\author{Enrique Arias\,\orcidicon{0000-0001-5064-8401}}
\email{email address: earias@iprj.uerj.br}
\affiliation{Universidade do Estado do Rio de Janeiro, 28625-570 Nova Friburgo, RJ, Brazil}

\author{G. O. Heymans\,\orcidicon{0000-0002-1650-4903 }}
\email{email address: olegario@cbpf.br}
\affiliation{Centro Brasileiro de Pesquisas F\'{\i}sicas, 22290-180 Rio de Janeiro, RJ, Brazil}
\author{H.~T.~Lopes}
\email{email address: herus@cbpf.br}
\author{N.~F.~Svaiter\,\orcidicon{0000-0001-8830-6925}}
\email{email address: nfuxsvai@cbpf.br}
\affiliation{Centro Brasileiro de Pesquisas F\'{\i}sicas, 22290-180 Rio de Janeiro, RJ, Brazil}


\begin{abstract}

We discuss the vacuum energy of a quantized scalar field in the presence of classical surfaces, defining bounded domains $\Omega \subset {\mathbb{R}}^{d}$, where the field satisfies ideal or non-ideal boundary conditions.
For the electromagnetic case, this situation describes the 
conductivity correction to the zero-point energy. 
Using an analytic regularization procedure, we obtain the vacuum energy for a massless scalar field 
at zero temperature in the presence of a slab geometry $\Omega=\mathbb R^{d-1}\times[0, L]$ with Dirichlet boundary conditions. To discuss the case of non-ideal boundary conditions, we employ an asymptotic expansion, based on an approximate functional equation for the Riemann zeta-function, where finite sums outside their original domain of convergence are defined.  
Finally, to obtain the Casimir energy
for a massless scalar field in the presence of a rectangular box, with lengths $L_{1}$ and $L_{2}$, i.e., $\Omega=[0,L_{1}]\times[0,L_{2}]$ with non-ideal boundary conditions, we employ an approximate functional equation of the Epstein zeta-function.

\end{abstract}


\pacs{05.20.-y, 75.10.Nr}

\maketitle


\section{Introduction}\label{intro}

Quantum fields are fundamental mathematical objects in 
the description of natural phenomena. These objects are operator-valued generalized functions with test functions, i.e., distributions in the Schwartz space \cite{Streater:1989vi,gelfand}. 
As a consequence, in  Minkowski spacetime, it has been shown that the renormalized vacuum expectation value of a quantum free scalar  field stress-energy tensor can exhibit a local negative energy density \cite{Epstein:1965zza}. In other words, although the energy operator associated with a quantum scalar field is self-adjoint and positive, the (0-0) component of the stress-energy tensor can be negative.
The Casimir effect is a measurable macroscopic manifestation of this result \cite{Casimir:1948dh,Ambjorn:1981xw,Fulling:1989nb,Plunien:1986ca, Bordag:2001qi, Milton:2004ya}. It has been measured in different geometric configurations \cite{Lamoreaux:1996wh,Mohideen:1998iz, capasso2007casimir, Klimchitskaya:2009cw}. The physical origin of the effect is the changes in the vacuum modes associated with the quantized electromagnetic field by the presence of macroscopic surfaces. The vacuum expectation values of the electric field at distinct points separated by space-like distances are correlated, like the interaction between atomic dipoles induced by the electromagnetic vacuum field (Van der Waals forces). Additionally, any constrained field, such as a massless fermionic field, can be a source of the effect as a consequence of the interaction of quantum field vacuum modes with idealized classical surfaces \cite{DePaola:1999im}. Another example is the phononic Casimir effect, where the speed of light is replaced by the speed of sound in the medium in the quasi-particle Landau scenario \cite{Ford:2009im}.

In the canonical formalism for bosonic and fermionic fields, the vacuum energy is divergent. To obtain a finite result, different approaches have been developed.   
One approach analyzes the local energy densities of quantized fields \cite{Brown:1969na,Bender:1976wb,Milton:1978sf,Kay:1978zr,Hays:1979bc,Actor:1994wv,Rodrigues:2003du}.
Another one, known as the global approach, investigates the total energy of the quantized field with idealized boundary conditions \cite{Fierz:1960zq, Boyer:1968uf,Boyer:1970bb}. 
This approach uses two natural ways to regularize and renormalize the divergent vacuum energy. The first one is the cut-off method, where an ultraviolet regulator function is introduced in the divergent sum of the eigenfrequencies. 
On general grounds, the regularized vacuum energy exhibits Weyl's terms with a geometric origin, cut-off independent contributions, and terms that vanish as the cut-off is removed. 
With these geometric terms in hand, we can implement a renormalization procedure with the introduction of auxiliary boundaries and subtraction of the regularized energies of different configurations. The second one are analytic regularization procedures. One is the spectral zeta-function regularization that was constructed to make sense of functional determinants. Using this procedure, the free energy of Euclidean quantum fields can be calculated \cite{Ray:1973sb, Hawking:1976ja, voros1992spectral}. 
Another analytic regularization procedure has also been discussed in \citep{Ruggiero:1977eu,Ruggiero:1979ec}. Although the cut-off method with the auxiliary configurations and the analytic regularization discussed above are quite different in their grounds, it is possible to compare them and shown to be analytically equivalent in some specific situations \cite{Svaiter:1989gz,Svaiter:1991je,Svaiter:1993nk,Svaiter:1994vd}. 

On the other hand, on physical grounds, the ideal boundary conditions of the perfect conductivity for all  electromagnetic field modes is an idealization. Usually, metallic plates behave as   
dielectric for high-frequency modes, and as conductors for infrared modes.
Following the original formulation, the question of the conductivity correction to the electromagnetic Casimir force arises.
To derive this correction, Lifshitz proposed a model treating the electromagnetic field  as a classical field, where attractive or repulsive forces arise from the fluctuating charges and currents of the boundaries \cite{Lifshitz:1956zz}.
 Further references include \cite{Schwinger:1977pa,Candelas:1981qw,lamoreaux1999calculation,Klimchitskaya:1999gg,bezerra2000higher,Lambrecht:1999vd,Torgerson:2004zz}.    

The purpose of this work is to discuss  the Casimir energy of a massless scalar field at zero temperature satisfying non-ideal boundary conditions. Due to the similarity between the quantized electromagnetic field and massless scalar fields satisfying Dirichlet and Neumann boundary conditions, our
problem has formal similarities with the conductivity correction to the Casimir force of the quantized electromagnetic field. Instead of discussing the non-linear problem of the microscopic modeling of finite conductivity i.e., non-ideal boundary conditions, we confine ourselves to make use of spectral theory of elliptic differential operators. This situation can be discussed using an analytic regularization procedure and approximate functional equations of spectral zeta-functions. These functional equations can express the Riemann and Epstein zeta-functions as finite sums outside their original domain of convergence. 

In our methodology, we use the fact that the total renormalized energy of scalar fields in the presence of bounded domains can always be derived using an analytic regularization procedure, where the Dirichlet and Neumann Laplacian are used. It's known that the vacuum energy in the slab geometry $\mathbb R^{d-1}\times[0, L]$ with Dirichlet boundary conditions can be written in terms of the Riemann zeta-function. 
To calculate its correction due to non-ideal boundary conditions, we represent the energy density using an asymptotic expansion derived by Hardy and Littlewood \cite{hardy1929approximate}. They obtained an
approximate functional equation for the Riemann zeta-function written as finite sums beyond their original domain of convergence. Next, we generalize the previous result in the case of a field in the presence of a rectangular box with lengths $L_{1}$ and $L_{2}$ with non-ideal boundary conditions. For instance, other generalizations of the Riemann functional equation have been presented in the literature. Recently it was discussed the introduction in the integral representation of the zeta-function, different cut-offs that are invariant under the transformation $x\mapsto 1/x$. It has been shown that the Riemann functional equation can be generalized with the same symmetry $s\to (1-s)$  in the critical strip \cite{Saldivar:2020dur}. 

This paper is organized as follows. 
In section \ref{sec:random} we discuss the  asymptotic behaviour of the eigenvalues of the Helmholtz wave equation, the 
Minakshisundaram-Pleijel zeta-function, the spectral decomposition of the heat kernel and
classical spectral invariants.
In section \ref{sec:randommass} we discuss how to obtain the renormalized vacuum energy for a massless scalar field at zero temperature in the presence of perfect mirrors. In section \ref{sec:quenchedfree} we use an approximate functional equation to obtain renormalized vacuum energy, due to non-ideal boundary conditions for a slab geometry $\mathbb R^{d-1}\times[0,L]$. 
In section \ref{sec:quenched} we employ the same method 
to obtain the renormalized vacuum energy for the case of a massless scalar field confined in a rectangular box, with lengths $L_{1}$ and $L_{2}$ with ideal and non-ideal boundary conditions.     
Conclusions are given in section \ref{sec:con}. Here we are using that $\hbar=c=k_{B}=1$. 

\section{Spectral properties of the Dirichlet Laplacian} \label{sec:random}

In this section, we want to describe briefly spectral methods which are fundamental tools to discuss problems in the definition of the global renormalized zero-point energy of scalar fields with ideal and non-ideal boundary conditions. Usually, to discuss vacuum energy issues and one-loop physics, it is necessary to introduce a normalization scale $\mu$. Since we are interested in discussing flat space-time that has boundaries and massless fields, the coefficient $c_{2}$ vanishes identically, and therefore the renormalized vacuum energy is independent of the normalization scale $\mu$. Consequently, we do not include the parameter $\mu$ in our equations. 

Consider the eigenfunctions and eigenvalues of the Laplacian operator $D=(-\Delta)$ on a bounded (open connected) domain $\Omega$ in Euclidean space 
$\mathbb R^{d}$. In this work, we discuss only the Dirichlet Laplacian which has a positive definite real spectrum. Also, the eigenvalues form a countable sequence. 
Using  $\lambda_{k}$ for $k=1,2...$ they are ordered as
\begin{equation}
0 < \lambda_{1}< \lambda_{2}\leq...
\leq\lambda_{k}\rightarrow \infty,
\end{equation}  
when $k\rightarrow\infty$, with possible multiplicities. The eigenfunctions  $\{\phi_{k}\}_{k=1}^{\infty}$ form a basis in the functional space $\mathcal{L}_{2}(\Omega)$ of measurable and square integrable functions on $\Omega$.  Each $\phi_{k}$ has eigenvalues $\lambda_{k}(\Omega)\equiv \lambda_{k}$.

In spectral theory, the asymptotic behaviour of the Dirichlet Laplacian eigenvalues in the analytic regularization procedure has a fundamental role.
This 
behaviour was investigated at first by Weyl \cite{weyl1912asymptotische}. Applying the Fredholm-Hilbert formalism of linear integral equations, it was proved that for $\Omega\subset \mathbb R^{d}$, $(d=2,3)$
\begin{equation}
\lim_{k\rightarrow\infty} \frac{k}{\lambda_{k}}=\frac{\mu_{d}(\Omega)}{4\pi},
\end{equation}
where $\mu_{d}(\Omega)$ is the Lebesgue measure of  
$\Omega$.

We start the discussion defining the density of eigenvalues as a sum of delta functions
\begin{equation}
g(\lambda)=\sum_{k}\delta(\lambda-\lambda_{k}),
\end{equation}
and the counting function $N(\lambda):=\#\{\lambda_{m}:\lambda_{m}<\lambda\}$, defined as 
\begin{equation}
N(\lambda)=\int_{0}^{\lambda }\mathrm{d}\lambda'g(\lambda'),
\end{equation}
which gives the number of elements in the sequence of eigenvalues, smaller than $\lambda$. The asymptotic behaviour of the counting function is given by
\begin{equation}
N(\lambda)=f(d)\mu_{d}(\Omega)\lambda^{\frac{d}{2}}, \,\,\,(\lambda\rightarrow \infty),
\end{equation}
where $f(d)$ is an entire function of $d$. 
Furthermore, the other asymptotic terms also give information about the boundary of the domain. As an example, for 
$\Omega\subset \mathbb R^{3}$, we get a contribution proportional to the surface area of $\Omega$.

Our first observation is about the zero-point energy renormalization. Let us define the Minakshisundaram-Pleijel bilocal zeta-function $\mathcal{Z}(x,y; s)$, for $s \in \mathbb C$ as
\begin{equation} 
\mathcal{Z}(x,y;s)= \sum_{k=1}^{\infty}\frac{\phi_{k}(x)\phi_{k}(y)}{\lambda_{k}^{s}}, 
\end{equation}
which converges uniformly in $x$ and $y$ for $Re(s)>s_{0}$ and was originally defined in a connected compact Riemannian manifold \cite{Minakshisundaram:1949xg}. From this bilocal zeta-function, is possible to define a spectral zeta-funciton associated with the eigenvalues of the Laplacian in $\Omega\subset \mathbb R^{d}$. We define $\mathsf{Z}(s) = Tr (-\Delta)^{-s}$, where
\begin{equation}
\mathsf{Z}(s) = \sum_{k=1}^\infty \lambda_k^{-s}
=\,\lim_{m\rightarrow\infty} \sum_{k=1}^{m}\lambda_k^{-s} .
\end{equation}
Using the counting function $N(\lambda)$ and the definition of the Riemann-Stieljes integral we get
\begin{eqnarray}
&\,&\sum_{n=1}^{m} \lambda_n^{-s} = \sum_{n=1}^{k-1} \lambda_n^{-s} + \int_a^b\mathrm{d}N(t )t^{-s}\,;   \\ &\,& \lambda_{k-1} \leq a < \lambda_k,\,\, \lambda_{m} \leq b <  \lambda_{m+1}.\nonumber
\end{eqnarray}
Therefore the spectral zeta-function can be written as 
\begin{equation}
\mathsf{Z}(s) = \sum_{n=1}^{k-1} \lambda_n^{-s} + \int_{\lambda_k}^{\infty} \mathrm{d}N(t)t^{-s} .
\end{equation}

In principle, this formula is given in the region of the complex plane where the original sum converges. As the sum on the right-hand side is analytic over the entire complex s-plane, the qualitative behavior of its analytic continuation is determined by the Riemann-Stieltjes integral expressed in terms of Weyl's counting function. 

To obtain the polar structure of the spectral zeta-function let us consider an evolution equation in $\mathcal{L}_{2}(\Omega)$ that can be formulated as the following initial-boundary problem in $(0,\infty)\times \Omega$. For $\Omega \subset \mathbb R^{d}$, we get 
\begin{equation*}
    \left\{\begin{aligned}
        &\frac{\partial u}{\partial t}=\Delta u\,,\nonumber\\
        &u(0,x)=f(x)\,,\nonumber\\
        &u(t,x)|_{x\in\partial\Omega}=0.
    \end{aligned}\right.
\end{equation*}
The weak solution $u(t,x)$, that  satisfies the diffusion equation in the sense of distributions is given by 
\begin{equation}
u(t,x)=\int \mathrm{d}\mu(y) p_{\Omega}(t,x,y)f(y),
\end{equation}
where $d\mu(y)$ is the volume element of the domain and
$p_{\Omega}(t,x,y)$ is the diffusion kernel, i.e., the positive fundamental solution to the heat equation. For a generic boundary condition, the spectral decomposition of the diffusion kernel can be represented as 
\begin{equation}
 p_{\Omega}(t,x,y)=\sum_{k=1}^{\infty}e^{-t\lambda_{k}(\Omega)}\phi_{k}(x)\phi_{k}(y).
\end{equation}
Using a Mellin transform and the definition of the Minakshisundaram-Pleijel zeta-function $\mathcal{Z}(x, y; s)$, we get  
\begin{equation}
\Gamma(s)\mathcal{Z}(x,y;s) = \int_0^{\infty} \mathrm{d}t \ t^{s-1} \  p_{\Omega}(t,x,y) . 
\end{equation}
For $x\neq y$, $\Gamma(s)\mathcal{Z}(x,y;s)$ is an regular function of $s$ in the entire complex plane. For $x=y$ there is a pole at $s=1$. From the diffusion kernel, since we are interested in global issues, let us define the trace of the diffusion kernel, written  as $\Theta(t) = Tr\left(e^{t\Delta}\right)$, where, using the Riemann-Stieljes integral we can write
\begin{equation}
\Theta(t)=\int_{0}^{\infty}e^{-\lambda t}\,dN(\lambda) = \sum_{k = 1}^\infty e^{-\lambda_k t}\,\,\,\,\,t>0.
\end{equation}
The spectral zeta-function can be represented as
\begin{equation} 
\mathsf{Z}(s) = \frac{1}{\Gamma(s)} \int_0^{\infty} \mathrm{d}t  \,t^{s-1} \Theta(t).
\end{equation}
Its polar structure in the extended complex plane is determined by the classical spectral invariants, which are the expansion coefficients at $t\rightarrow 0^{+}$ of the diffusion kernel trace. 
When $\partial\Omega\neq \varnothing $ the coefficients of the asymptotic expansion of the heat trace have been calculated for a variety of boundary conditions
 
\begin{equation}
\lim_{t \rightarrow 0^+}\Theta(t)=(4\pi t)^{-\frac{d}{2}}\Biggl[\,\sum_{p=0}^{K}c_{p}(\Omega)t^{\frac{p}{2}}
+o(t^{\frac{K+1}{2}})\Biggr],
\end{equation}
where the coefficients $c_{p}(\Omega)$ are related to the geometric characteristics of the bounded domain. By a Tauberian theorem, we are able to connect the first term of the above asymptotic expansion with Weyl's asymptotic behavior of the Laplace operator spectrum.

For the case of vacuum energy, Fulling has stressed the need to study the cylinder kernel 
\cite{Fulling:2003zx,Fulling:2014poa}. See, for example, \cite{bar2003heat}.
To implement this idea, let us define the zeta-function $\zeta_{\sqrt{D}}(s)$  
constructed with the energies $\omega_{k}$ of each normal modes 
\begin{equation}
\zeta_{\sqrt{D}}(s)=\sum_{k=1}^\infty \frac{1}{\omega_k^{s}}\,; \qquad Re(s) > s_1. 
\end{equation}
The renormalized vacuum energy is by definition $\langle E\rangle_{r}=\zeta_{\sqrt{D}}(s)|_{s=-1}$. Using again a Mellin transform 
we have 
\begin{equation}
\sum_{k=1}^\infty \frac{1}{\omega_k^{s}} = \frac{1}{\Gamma(\frac{s}{2})} \int_0^{\infty} \mathrm{d}t \ t^{\frac{s}{2}-1} \  \sum_{k=1}^\infty e^{-\omega^2_k t}.
\end{equation}
The zeta-function $\zeta_{\sqrt{D}}(s)$ is a meromorphic function of $s$ with simple poles. In the case where $s=-1$ is a pole, we can obtain a representation in a neighborhood of the pole, including some regular part known as the renormalized vacuum energy. It is important to stress that the measurable Casimir energy is obtained from a mathematical formalism based on analytic continuations, where undesirable polar contributions must be removed through a renormalization procedure. 

\section{The  vacuum energy in the presence of surfaces with ideal boundary conditions} \label{sec:randommass}

The aim of this section is to use an 
analytic regularization procedure to obtain the 
vacuum energy of a massless scalar field 
at zero temperature in the presence a slab geometry $\Omega=\mathbb R^{d-1}\times[0,L]$ with Dirichlet boundary conditions. Let us assume a free neutral scalar field defined in a 
$(d+1)$-dimensional flat space-time. 
Its field equation, the Klein-Gordon equation, reads
\begin{equation}
\left(\frac{\partial^{2}}{\partial t^{2}}-\Delta+m_{0}^{2}\right)\varphi(t,\x)=0.
\end{equation}
To implement the canonical quantization, the field operator and the generalized momentum are expanded in  Fourier series enclosed in a finite periodic box. With a defined operator Hamiltonian, $\mathcal{H}$, the energy of the confined field has a pure point spectrum, allowing us to characterize its states in terms of occupation numbers and the Fock representation. For Dirichlet boundary conditions, the situation is similar.
The states of the system are described in terms of occupation numbers of elementary excitations, which characterizes the states concerning the ground state, the vacuum state $|0\rangle$.

To proceed, we restrict the field to a $d$-dimensional box with lengths $(L_{1}\times L_{2}\times...\times L_{d-1}\times L_{d})$. Assuming Dirichlet boundary conditions, the vacuum energy, i.e., the total energy of the quantized field in the box, is $\langle 0|\mathcal{H}|0\rangle=U_{d}(L_{1},...,L_{d-1},L_{d})$.  Using the condition $L_{d}\ll L_{i}$  for $(i=1,2,...,d-1)$, and defining $L_{d}=L$, the unrenormalized vacuum energy can be written as   
\begin{eqnarray}
&\,&U_{d}(L_{1},...,L_{d-1},L)=\frac{1}{(2 \pi)^{d-1}}\left(\prod_{i=1}^{d-1}L_{i}\right)\nonumber\\
&\times& \int\prod_{i=1}^{d-1}\mathrm{d}q_{i}\sum_{n=1}^{\infty}\left(q_{1}^{2}+...+q_{d-1}^{2}
+\left(\frac{n\pi }{L}\right)^{2}+m_{0}^{2}
\right)^{\frac{1}{2}}.\nonumber \\
\end{eqnarray}
To discuss the case similar to the electromagnetic field let us assume $m_{0}^{2}=0$. The unrenormalized vacuum energy per unit area is defined as  
\begin{equation}\label{eq:density}
\epsilon_{d}(L)=\frac{U_{d}(L_{1},...,L_{d-1},L)}{\left(\prod_{i=1}^{d-1}L_{i}\right)},
\end{equation}
and this is a divergent expression. It can be written as
\begin{equation}
\epsilon_{d}(L)=\frac{\left(4\pi\right)^{\frac{1-d}{2}}}{\Gamma\left(\frac{d-1}{2}\right)} \sum_{n=1}^{\infty}\int_{0}^{\infty} \mathrm{d}r\, r^{d-2}\left[r^{2}+\left(\frac{n\pi}{L}\right)^{2}\right]^{\frac{1}{2}}.
\label{imp}
\end{equation}
%
%
%
A straightforward  calculation  
led us to 
\begin{equation}
\epsilon_{d}(L)=\frac{\left(4\pi\right)^{\frac{1-d}{2}}}{2\Gamma\left(\frac{d-1}{2}\right)}
\left(\frac{\pi}{L}\right)^{d}
\int_{0}^{\infty} \mathrm{d}x\, x^{\frac{d-3}{2}}(1+x)^{\frac{1}{2}}
\sum_{n=1}^{\infty}n^{d}.
\end{equation}
In the limit $L\rightarrow \infty$ we should obtain the fundamental result that the vacuum is a Lorentz invariant state of zero energy.
Using the definition of the Beta function as 
\begin{equation}
\mathcal{B}(z,w)=\frac{\Gamma(z)\Gamma(w)}{\Gamma(z+w)}\,,
\end{equation} 
and an analytic continuation principle, the vacuum energy per unit area is given by
\begin{equation}\label{eq:casimir}
\epsilon_{d}(L)=-\frac{\pi^{\frac{d}{2}}\Gamma\left(-\frac{d}{2}\right)}{2(2L)^{d}}\zeta(-d),
\end{equation}
%
%
where $\zeta(s)$ is the Riemann zeta-function,  which is a function of the complex variable $s=\sigma+i t$, where $\sigma, t\,\, \in{\mathbb{R}}$. It is originally defined in the half-plane $\Re\,(s)>1$
through an absolutely convergent Dirichlet series \cite{riemann1859ueber,ingham1990distribution}. The series is defined by summing over the set of natural numbers $n\in \mathbb{N}$ and can be expressed as
\begin{equation}
\zeta(s)= \sum_{n=1}^{\infty}\,\frac{1}{n^{s}}.
\label{p2}
\end{equation}
It can be extended to the complex plane as a meromorphic function using the Poisson summation formula with a simple pole at $s=1$.

It is possible to show that Riemann zeta-function $\zeta(s)$ satisfies a functional equation
valid for $s\in {\mathbb{C}} \setminus\left\{0,1\right\}$. This equation connects two functions outside the original domain of convergence. 
Using the properties of the Gamma function to define $\vartheta(s)$ as
\begin{equation}\label{ef}
\vartheta(s)= \frac{(2\pi)^{s}\Gamma\left(1-s\right)}{\Gamma\left(1 -\frac{s}{2} \right) \Gamma\left(\frac{s}{2}\right)} ,
\end{equation}
we get a reflection formula for the Riemann zeta-function
\begin{equation}
    \zeta(s) = \vartheta(s)\zeta(1-s).
\end{equation}

The above calculations are an intermediate step crucial to discuss the modifications in the renormalized vacuum energy of a scalar field in the presence of surfaces where the scalar field satisfies non-ideal boundary conditions.

\section{Renormalized Vacuum energy with Non-ideal boundary conditions} \label{sec:quenchedfree}

In the Lifshitz approach, the dispersion forces between dissipative media are caused by the fluctuating electromagnetic field defined both within and outside the media. Using the fluctuation-dissipation theorem,
the Lifshitz expression for the force between plates depends on the dielectric functions on the surfaces and also on the medium in which they are immersed. The finite conductivity correction to the ideal Casimir calculation is obtained using the frequency dependence of the dielectric function.  
The imperfect conductivity at high frequencies can be modeled by introducing only the plasma frequency $\omega_{p}$ of the plates. It is  important to note that the Casimir result is recovered at distances larger than the plasma wavelength.  

In our case, we are discussing the vacuum energy of a quantized scalar field in the presence of classical surfaces, where the field satisfies non-ideal boundary conditions. Those can be understood as finite conductivity conditions. However, the crucial point is that is not convenient to simply calculate the correction to the renormalized vacuum energy separating the effects of the low-energy vacuum modes from the high-energy modes using a sharp cut-off, once that is a sum of positive terms one always obtain a positive energy density, i.e.,
\begin{equation}
    \epsilon_{d}^{\mathrm{f.c.}}(L) = \sum_{k = 1}^{k_c} \omega_k > 0,
\end{equation}
where $\omega_{k_c + 1}$ is plasma frequence of the material.

We start using an analytic regularization procedure and the fact that for Dirichlet boundary conditions the eigenvalues vary continuously under a smooth deformation of the domain (spectral stability of elliptic operator under domain deformation) and the minimax principle says that the eigenvalues monotonously decrease when the domain is enlarger,
\begin{equation}
\sigma_{m}(\Omega_{1})\geq \sigma_{m}(\Omega_{2}),\,\,\, \Omega_{1}\subset \Omega_{2}.
\end{equation}
By the above arguments, we can use 
approximate functional equation that expresses the Riemann zeta-function as finite sums, outside their original domain of convergence.

Initially, we use a classical result by Hardy and Littlewood following the derivation discussed in Ref. \cite{ivic2012riemann}. 
Let us write the Riemann zeta-function as

\begin{eqnarray}
\zeta(s)&=&\sum_{n\leq n_{c}}n^{-s}+\sum_{n>n_{c}}n^{-s}\nonumber\\
&=&\sum_{n\leq n_{c}}n^{-s}+\frac{1}{\Gamma(s)}
\int_{0}^{\infty}\mathrm{d}x\,x^{s-1}\left(\sum_{n>n_{c}}e^{-nx}
\right)\nonumber\\
&=&\sum_{n\leq n_{c}}n^{-s}+\frac{1}{\Gamma(s)}
\int_{0}^{\infty}\mathrm{d}x\,\frac{x^{s-1}\,e^{-n_{c}x}}{e^{x}-1},
\end{eqnarray}
where  the absolute convergence justifies the inversion of the order of summation and integration. To proceed, we analyze the following integral $I(s)$. We have
\begin{equation}
I(s)=\int_{C}\mathrm{d}z\,\frac{z^{s-1}\,e^{-n_{c}z}}{e^{z}-1},
\end{equation}
where the contour $C$ starts at infinity on the positive real axis, encircles the origin once in the positive direction excluding the points $\pm 2\pi i,\pm 4\pi i,...$ and returns to infinity. We obtain
\begin{equation}
I(s)=\left(e^{2\pi\, i s}-1\right)\int_{0}^{\infty}\mathrm{d}x\,\frac{x^{s-1}\,e^{-n_{c}x}}{e^{x}-1}.
\end{equation}
Using the analytic continuation principle we can write
\begin{equation}
\zeta(s)=\sum_{n\leq n_{c}}n^{-s}+\frac{e^{-\pi i s}\Gamma(1-s)}{2\pi i}\int_{C}\mathrm{d}z\,\frac{z^{s-1}\,e^{-n_{c}z}}{e^{z}-1}.
\end{equation}
From the above equation, we find an approximate representation of the zeta-function in terms of finite sums. It was proved that
\begin{equation}\label{eq:hardy}
\zeta(s)=\sum_{n\leq\,x}\!\frac{1}{n^{s}}+\vartheta(s)\sum_{n\leq\,y}\frac{1}{n^{1-s}}\!+\!O(x^{-\sigma})+O(t^{\frac{1}{2}-\sigma}\,y^{\sigma-1}),
\end{equation}
for $0\leq\,\sigma<1$ holds for given $x,y,t >C>0$ satisfying $2\pi xy=t$ where $t \gg 1$. This is known as an approximate functional equation. 

For simplicity, using the approximate functional equation, we discuss the case of a slab geometry $\mathbb R^{d-1}\times[0,L]$. Making a parallel with the electromagnetic case, in the scalar field scenario, we define the plasma frequency $\omega_{p}$ and the plasma wavelength $\lambda_{p}=2\pi/w_{p}$. Next, we define a ``critical" mode index $n_{c}$, which will be related to the plasma wavelength.
In order to find an adequate maximum number of states $n_c$ for a single compactified direction, we need to introduce first the notion of density of states $\rho(k)$ in the phase space and the number of states $dN=\rho(k)d^dk$ that lies between $k$ and $k+dk$. 
In our $d$-dimensional space, where all the directions are finite and have lengths $L_1, L_2,...,L_{d-1},L$, then the density of states is simply 
\begin{equation}
\rho(k)=\left(\frac{L}{\pi^d}\right)\prod_{i=1}^{d-1}L_i \,,   
\end{equation}
we can find the number of states inside a volume that possess the maximum value of moment $k_{max}$ as
\begin{equation}
N(k_{max})=\int_{|k|<k_{max}} \mathrm{d}^d k\rho(k)=\rho\,\,\frac{\pi^{d/2}}{\Gamma(\frac{d}{2}+1)}k_{max}^d\,,    
\end{equation}
where we have used the definitions of the volume of a sphere in $d$-dimensions. For other side, we are interested in obtaining the maximum number of states in a single compactified direction $n_c$. We have that
\begin{equation}
N(k_{max})=\frac{\pi^{d/2}}{\Gamma(\frac{d}{2}+1)}n_c^d.    
\end{equation}
Therefore we identified 
$n_c^d=\rho k_{max}^d$.
Now, we relate the maximum wave number with the plasma frequency of the material in such a manner that $k_{max}=2\pi/\lambda_p$. With all this, after some algebra, we conclude that
\begin{equation}
n_c=2\left(\frac{L^{1/d}}{\lambda_p}\right)\prod_{i=1}^{d-1}L_i^{1/d},    
\end{equation}
since all the directions $L_i$ from $i=\{1,2,...,d-1\}$ are much larger that $L$. The only dependence of the maximum number of states is of the form
\begin{equation}
    n_{c}(L)\equiv \left(\frac{L}{\lambda_{p}}\right)^{1/d}\,\,.
\end{equation}
In the Hardy and Littlewood approximate functional equation, we choose
\begin{equation}
    x=y=\left(\frac{L}{\lambda_{p}}\right)^{1/d} = n_c \quad \Rightarrow \quad t=2\pi \left(\frac{L}{\lambda_{p}}\right)^{2/d} = 2\pi n_c^2\,\,.
\end{equation}
Using the asymptotic expansion, Eq. (\ref{eq:hardy}), we get the Casimir energy as 
\begin{equation}\label{eq:H-L}
\epsilon_{d}(L)=-\frac{\pi^{\frac{d}{2}}\Gamma\left(-\frac{d}{2}\right)}{2(2L)^{d}} \left[ H_{n_{c}}(-d)+\vartheta(-d)H_{n_{c}}(d+1)\right].
\end{equation}
The quantities $H_n(s)$ are the generalized harmonic numbers. Once the Eq. (\ref{eq:H-L}) only makes sense as an analytic continuation, those finite sums must be understood as such. 
Moreover, we stress the fact that the equality holds by analytic continuation outside the strip $0<\sigma<1$.
This can be shown using an analytic continuation of the asymptotic expansion. 

Each generalized harmonic number has an expression for its domain of interest in the complex plane. Lets us start from the second term in the sum, $H_{n_c}(d+1)$. Formally, this quantity is given by
\begin{equation}
    H_{n_c}(d+1) \equiv \sum_{n=1}^{n_c}\frac{1}{n^{d+1}}.
\end{equation}
However, since we start from Eq. (\ref{eq:casimir}), which is an analytic continuation, the finite sum should be taken in the range of interest. In such a situation, we can use a known expression
\begin{equation}
    H_{n_c}(d+1) = \zeta(d+1) + \frac{(-1)^{d}}{d!}\psi_{d}(n_c+1),
\end{equation}
which holds for $n_c\in \mathbb{R}\setminus\{-1,-2,-3,\dots\}$ and $d\in\mathbb{N}$; see e.g. \cite{sofo2018general}, and $\psi_m(x)$ is the polygamma function. Using a recurrence relation and a expression for large arguments, we can write the polygamma function as
\begin{equation}
    \psi_{d}(n_c+1) = \frac{(-1)^d d!}{n_c^{d+1}}+(-1)^{d+1}\sum_{k=0}^{\infty} \frac{(k+d-1)!}{k!}\frac{B_k}{n_c^{d+k}}\,,
\end{equation}
where $B_k$ are the Bernoulli numbers (we take the convention $B_1=1/2$). Using the definition of $n_c$ and in the limit of $L/\lambda_p \gg 1$ we can write
\begin{equation}
    \psi_{d}(n_c+1)\approx (-1)^{d+1}\left(\frac{\lambda_p}{L}\right)\left[(d-1)! - \frac{1}{2}d!\left(\frac{\lambda_p}{L}\right)^{\frac{1}{d}}\right],
\end{equation}
which allow us to write the $H_{n_c}(d+1)$ in powers of $\lambda_p/L$.

For the first term of Eq.(\ref{eq:H-L}), we formally have
\begin{equation}
    H_{n_c}(-d) \equiv \sum_{n=1}^{n_c}\frac{1}{n^{-d}},
\end{equation}
and an analytic continuation can be obtained using some elementary operations and the uniqueness of the analytic continuation, is straightforward to see that
\begin{equation}\label{eq:GHNH}
    H_{n_c}(-d) = \zeta(-d) - \zeta_H(-d;n_c+1),
\end{equation}
where $\zeta_H(-d;n_c+1)$ is the Hurwitz zeta-function, defined by
\begin{equation}
    \zeta_H(s;a) \equiv \sum_{n=0}^{\infty}\frac{1}{(n+a)^s}.
\end{equation}

Let us define the Casimir energy per unit area with non-ideal boundary conditions, i.e., finite conductivity (f.c.) by 
\begin{equation}
    \epsilon^{\textrm{f.c.}}_{d}(L) \equiv -\frac{1}{L^d}\frac{\pi^{d/2}}{2^{d+1}}\Gamma\left(-\frac{d}{2}\right)\zeta_H(-d;n_c+1).
\end{equation}
Once this is performed, we can identify the contribution from the ideal boundary conditions, and the remaining part can be regarded as a correction term. We get

\begin{eqnarray}
    \epsilon^{\textrm{f.c.}}_{d}(L) &=&  \epsilon_{d}(L)  \nonumber \\ &+& \frac{\Gamma(1+d)\lambda_p}{2 \ \Gamma\left(1+\frac{d}{2}\right)}\left(\frac{1}{4\sqrt{\pi}}\right)^d 
    \left[\frac{1}{L^{d+1}d} - \frac{\lambda_p^{\frac{1}{d}}}{2 L^{d+1+\frac{1}{d}}}\right].\nonumber \\
\end{eqnarray}

As we have observed, in the slab geometry, the Casimir force is a negative quantity ($\epsilon_{d}(L) < 0$), while the second contribution in the above equation is positive. We have succeeded in deriving the Casimir energy per unit area with non-ideal boundary conditions. Note that our first finite conductivity correction to the electromagnetic Casimir energy is the same as the correction obtained using the Lifshitz calculations. In contrast, the second correction is smaller, with the Lifshitz formula giving a second correction as $L^{-5}$, whereas ours give $L^{-\frac{13}{3}}$. The basic assumption that needs to be carefully investigated is the discussion of vacuum energy in a bounded domain. To proceed, in the next section, we generalize the above result to the $d=2$ dimensional case for a finite volume box.

\section{Casimir energy in a rectangular box  with non-ideal boundary conditions} \label{sec:quenched}

Let us discuss now the eigenvalues of a second-order elliptic self-adjoint partial differential operator on scalar functions on a bounded domain. We consider the eigenvalues of 
$-\Delta$ on a connected open set $\Omega$ in Euclidean space $\mathbb{R}^{2}$. We assume that the massless scalar field is confined in a rectangular box, with lengths $L_{1}$ and $L_{2}$ obeying Dirichlet boundary conditions. The eigenfrequencies that we use to expand the field operator are given by
\begin{equation}
\omega_{n_1 n_2}=\Biggl[\Bigl(\frac{n_1\pi}{L_{1}}\Bigr)^{2}+
\Bigl(\frac{n_2\pi}{L_{2}}\Bigr)^{2}
\Biggr]^{\frac{1}{2}};\,\,\,\,n_1,n_2=1,2,....
\end{equation}
The unrenormalized vacuum energy in this case is
\begin{equation}
U(L_{1}, L_{2})=\frac{1}{2}\sum_{n_1,n_2=1}^{\infty}\omega_{n_1n_2}.
\end{equation}
Making use of an analytic regularization procedure, the divergent expression can be written as 
\begin{equation}
E(L_{1}, L_{2},s)=\frac{1}{2}\sum_{n_1,n_2=1}^{\infty}\omega_{n_1n_2}^{-2s},
\end{equation}
for  $s\in\mathbb{C}$. Observe that, the vacuum energy is obtained when $s=-\frac{1}{2}$. The above double series converges absolutely and uniformly for $Re(s)>1$. 
An  analytic function, which plays an important role in algebraic number theory is the 
Epstein zeta-functions associated with quadratic forms \cite{epstein1903theorie}. Suppose that
\begin{equation}
\phi(a,b,c; x,y)=ax^{2}+cxy+by^{2},
\end{equation}
where $a,b$ and $c \in \mathbb{R}$ and $a>0$ and $\eta=4ab-c^{2}>0$. Lets us define the function $\mathcal{A}(s)$ by the series
\begin{equation}
\mathcal{A}(a,b,c;s)=\sideset{}{'}\sum_{n_1,n_2=-\infty}^{\infty}
\phi^{-s}(a,b,c; n_{1},n_{2}),
\end{equation}
The above series defines an analytic function for  $s=\sigma+it$, ($\sigma\in \mathbb{R}$ and $t\in \mathbb{R}$) and $\sigma >1$, where we adopt the notation that the prime sign in the summation means that the contribution $n_1=n_2=0$ (the origin of the mode space) must be excluded. This particular case of the Epstein zeta-function can be continued analytically to the whole complex plane, except for a simple pole at $s=1$ \cite{Ford:1994gx}. This double series exhibits a functional equation that can be obtained using properties of the theta-function or the Poisson summation formula. The functional equation reads
\begin{equation}
\mathcal{A}(a,b,c;s)=\left(\frac{2\pi}{\sqrt{\eta}}\right)^{2s-1}
\frac{\Gamma(1-s)}{\Gamma(s)}\mathcal{A}\left(\frac{1}{a},\frac{1}{b},\frac{1}{c};1-s\right)
\end{equation}
We are interested in the case where $c=0$. Let us define the function
$Z\left(\frac{1}{L_{1}},\frac{1}{L_{2}};s\right)$ by
\begin{equation}
Z\left(\frac{1}{L_{1}},\frac{1}{L_{2}};s\right)=\sideset{}{'}\sum_{n_1,n_2=-\infty}^{\infty}
\left(\frac{n_{1}^{2}}{L_{1}}+\frac{n_{2}^{2}}{L_{2}}\right)^{-s},
\end{equation}
We can find that the vacuum energy is written as 
\begin{eqnarray}\label{eq:CasBox}
E(L_{1}, L_{2};s)&=&\frac{1}{8}Z\left(\frac{\pi^2}{L_{1}^2},\frac{\pi^2}{L_{2}^2};s\right)\nonumber\\&-&\frac{1}{4}\left[\left(\frac{\pi}{L_{1}}\right)^{-2s}+\left(\frac{\pi}{L_{2}}\right)^{-2s}\right]\zeta(2s). \nonumber \\
\end{eqnarray}

As it was discussed, $E(L_{1}, L_{2},s)$ is analytic in 
 $s\in\mathbb{C}\setminus \lbrace\frac{1}{2},1\rbrace$. Using the analytic continuation of the Epstein and the Riemann zeta-function gives the vacuum energy  $U(L_{1},L_{2})=E(L_{1}, L_{2};s=-1/2)$ for the system with Dirichlet boundary conditions. 
We get
\begin{eqnarray} 
U(L_{1},L_{2})&=&\frac{\pi}{48}\left(\frac{1}{L_{1}}+\frac{1}{L_{2}}\right)\nonumber\\
&-& \frac{L_{1}L_{2}}{32\pi}\sideset{}{'}\sum_{n_1,n_2=-\infty}^{\infty}\left(
n_1^{2}L_{1}^{2}+n_2^{2}L_{2}^{2}\right)^{-\frac{3}{2}}.\nonumber \\
\end{eqnarray}

The next step involves discussing the scalar case similar to the electromagnetic case of imperfect conductors, where there is a plasma frequency $\omega_{p}$. Using the same approach discussed in the previous section, we aim to determine the approximate functional equation for the Epstein zeta-function.

Potter \cite{potter1934approximate} has derived the following approximate functional equation:
\begin{eqnarray}
    \mathcal{A}(a,b,c;s) &=& \sideset{}{'}\sum_{\phi \leq x} \phi^{-s}(a,b,c;n_1,n_2) \nonumber \\
    &+& X(s)\sideset{}{'}\sum_{\phi \leq y} \phi^{s-1}(a,b,c;n_1,n_2) ,
\end{eqnarray}
for $t \gg 1$, and the condition $4\pi^2 xy =\eta \,t^2 $ must be satisfied, the quantity $X(s)$ is defined by
\begin{equation}
    X(s) = \left(\frac{2\pi}{\sqrt{\eta}}\right)^{2s-1}\frac{\Gamma(1-s)}{\Gamma(s)}.
\end{equation}
Henceforth we take $\mathcal{A}(a,b,0;s)\equiv \mathcal{A}(a,b;s)$ and similar for $\phi$.

Of course, to obtain the correction to the Casimir energy via asymptotic series we will need to use the Potter approximate functional equation for the Epstein zeta function, but also the Hatree-Littlewood approximate functional equation for the Riemann zeta-function. Let's start analyzing the Epstein zeta-function. It's convenient to introduce a $\lambda_p$ term in our expression in order to  only have adimensional quantities and establish a parallel with the Casimir energy in a finite conductivity scenario. In this case, we have

\begin{equation}\label{eq:potter}
    \mathcal{A}\left(\frac{\pi^2\lambda_p^2}{L^2_1} , \frac{\pi^2\lambda_p^2}{L^2_2};s\right) = \sideset{}{'}\sum_{\Phi \leq x} \Phi_{12}^{-s} + X(s)\sideset{}{'}\sum_{\Phi \leq y} \Phi_{12}^{s-1}\,,
\end{equation}
where to the notation be lightened, we defined
\begin{eqnarray}
    \Phi_{12} &\equiv& \phi\left(\frac{\pi^2\lambda_p^2}{L^2_1} , \frac{\pi^2\lambda_p^2}{L^2_2};n_1,n_2\right) \nonumber \\
    &=& \frac{\pi^2\lambda_p^2}{L^2_1} n_1^2 +\frac{\pi^2\lambda_p^2}{L^2_2} n_2^2 \,,
\end{eqnarray}
once that $4\pi^2 xy = \eta \, t^2$ with
\begin{eqnarray}
    \eta = 4\left(\frac{\pi^2\lambda_p^2}{L_1L_2}\right)^2 \Rightarrow xy = \left(\frac{\pi\lambda_p^2}{L_1L_2}\right)^2t^2.
\end{eqnarray}
Since 
\begin{equation}
   X(s) = \left(\frac{L_1L_ 2}{\pi \lambda_p^2}\right)^{2s-1}\frac{\Gamma(1-s)}{\Gamma(s)},
\end{equation}
 using a similar argument that we used before, but now all dimensions remain compact, we can define the quantities 
\begin{eqnarray}
    n_c^{(1)} &\equiv& \left(\frac{L_1}{\lambda_p}\right)^{1/2} \,\,\, \mathrm{and} \,\,\, n_c^{(2)} \equiv \left(\frac{L_2} {\lambda_p}\right)^{1/2} \nonumber \\ 
    &\Rightarrow& xy = \left[\frac{\pi}{\left(n_c^{(1)} n_c^{(2)}\right)^2}\right]^2 t^2 ,
\end{eqnarray}
which, considering the fact that we do not have a preferred direction, indicate to us that the natural choice for $t$ should be 
\begin{equation}
    t = \frac{1}{\pi}\left(n_c^{(1)} n_c^{(2)}\right)^2 \Rightarrow x = y = n_c^{(1)} n_c^{(2)}.
\end{equation}
So looking back to the Eq. (\ref{eq:potter}), we see that the sums are over all modes inside the ellipse defined by 
\begin{equation}
\frac{n_1^2}{L_1n_c^{(1)} n_c^{(2)}} + \frac{n_2^2}{L_2 n_c^{(1)} n_c^{(2)}} = \left(\frac{1}{\pi \lambda_p}\right)^2 =\,\, \textrm{constant},
\end{equation}
in the $(n_1,n_2)$-plane with the origin removed.

For the Riemann zeta-function contributions, that are present in Eq. (\ref{eq:CasBox}), we have
\begin{equation}
    \zeta(2s) = \sum_{n \leq u}\frac{1}{n^{2s}} + \vartheta(2s)\sum_{n \leq v} \frac{1}{n^{1-2s}},
\end{equation}
for $\alpha \gg1$ where $2\pi uv = \alpha$. Proceeding exactly as in the slab bag geometry case, we find that
\begin{equation}
    u = v \equiv  n_c^{(i)} = \left(\frac{L_i}{\lambda_p}\right)^{1/2} \,\,  \Rightarrow  t=2\pi \frac{L_i}{\lambda_p}; \quad i = 1, \ 2,
\end{equation}
continuing from the previous section, we employ an analogous method. Using the same harmonic number definitions, once the range in the complex plane will be the same. Considering the case where $s=-1/2$ and manipulating the equations, is possible to find that.

\begin{widetext}
\begin{eqnarray}
E(L_1,L_2; s) &=& \frac{\lambda_p^{2s}}{8} \sideset{}{'}\sum_{\Phi \leq n_c^{(1)}n_c^{(2)}} \Phi_{12}^{-s}+ \left(\frac{L_1L_ 2}{\pi \lambda_p^2}\right)^{2s-1}\frac{\Gamma(1-s)}{\Gamma(s)}\frac{\lambda_p^{2s}}{8}\sideset{}{'}\sum_{\Phi \leq n_c^{(1)}n_c^{(2)}}\Phi_{12}^{s-1}  \nonumber 
\\ - \frac{\lambda_p^{2s}}{4} \sum_{i=1}^2\Biggl\{\left(\frac{\lambda_p}{L_i}\right)^{-2s}\Biggl[2\zeta(2s) 
&-&\left. \zeta_H(2s;n_c^{(i)}+1)\Biggr] + (-1)^{-4s +1}\vartheta(2s)\left[\frac{1}{2s}\left(\frac{\lambda_p}{L_i}\right)^{-3s} -\frac{1}{2} \left(\frac{\lambda_p}{L_i}\right)^{\frac{-6s+1}{2}}\right] \right\}.
\end{eqnarray}


We define the vacuum energy for finite conductivity (f.c.) as 

\begin{equation}
 E^{\textrm{f.c.}}\left(L_{1}, L_{2},s = -\frac{1}{2}\right) = U^{\textrm{f.c.}}(L_1, L_2)  \equiv \frac{1}{8\lambda_p}\sum_{\Phi \leq n_c^{(1)} n_c^{(2)}}  \Phi_{12}^{\frac{1}{2}} - \frac{1}{4}\sum_{i=1}^2\frac{1}{L_i}\left[\zeta_H(-1;n_c^{(i)}+1) - \frac{1}{6}\right]. 
\end{equation}

Therefore

\begin{equation}
    U^{\textrm{f.c.}}(L_1, L_2) = U(L_1,L_2) - \frac{\pi^2\lambda_p^3}{32 (L_1 L_2)^2} \sum_{\Phi \leq n_c^{(1)} n_c^{(2)}}\Phi_{12}^{-\frac{3}{2}} +\frac{1}{2\lambda_p(2\pi)^2}\sum_{i=1}^2\left[\left(\frac{\lambda_p}{L_i}\right)^{3/2} -  \frac{1}{2}\left(\frac{\lambda_p}{L_i}\right)^{2}\right],
\end{equation}
is the Casimir energy for a rectangular box with non-ideal boundary conditions.
\end{widetext}
\newpage
\section{Conclusions} \label{sec:con}

In this paper, we investigate the total energy of a  massless scalar quantized field, which satisfies idealized perfectly boundary conditions, using an analytic regularization procedure. We extend the above result to the case of ``imperfect conductor" boundary conditions. The crucial point  in this scenario is that it is not convenient to calculate the correction to the renormalized vacuum energy  
separating the effects of the low-energy vacuum modes from the high-energy modes using a cut-off method without realizing previously a regularization of the zero-point energy. 
Therefore, to obtain the correction to the Casimir force for imperfect conductors assuming a slab geometry $\mathbb R^{d-1}\times[0,L]$, we have to use an approximate functional equation. First we represent the energy density using finite sums outside the original domain of convergence of the Dirichlet series. Next, we demonstrate how it is possible to obtain the correction to the force in the three-dimensional spacetime generated by a massless scalar field in the presence of a rectangular box, with lengths $L_{1}$ and $L_{2}$. 

In the literature, it has been discussed a scenario where classical fluctuations assume the role of the quantum vacuum modes as the original Casimir conceptual framework
\cite{fisher2003phenomenes,krech1999fluctuation,Kardar:1997cu,brankov2000theory,gambassi2009casimir,dohm2013crossover,Gross:2017mvq, dantchev2023critical}. 
In a confined system with quenched disorder, a sensitivity to the boundaries may arise, where the distance to the critical situation is given by some non-thermal control parameter. 
Recently, inspired by the statistical Casimir effect, it was discussed the application of the spectral and distributional zeta-function methods to describe fluctuation-induced forces arising from a quenched disorder field in a continuous Landau-Ginzburg model \cite{Svaiter:2016lha}. 
A series of representations was employed. 
From the series representation of the average free energy, it is possible to obtain the force between the boundaries, due to the interaction of the critical fluctuations generated by the moment of the partition function, with the largest correlation length of the fluctuations.  
In other words, varying continuously the intensity of a non-thermal control parameter, the induced force can be repulsive or attractive between the boundaries \cite{Rodriguez-Camargo:2022wyz}. This is the problem of the sign of the Casimir force in the statistical Casimir effect \cite{Bimonte:2014ovq}. 
It is clear that is possible to go beyond the Gaussian approximation, where a perturbative expansion must be implemented with Euclidean Green's functions. In this case, in addition to the traditional bulk counterterms, surface counterterms must be introduced to renormalize the interacting Euclidean field theory in the presence of boundaries \cite{Symanzik:1981wd,Fosco:1999rs,Caicedo:2002ft,AparicioAlcalde:2005wxe}. Therefore, a natural continuation of this work  would involve investigating the statistical Casimir effect in the presence of dirty surfaces. \cite{diehl1990critical}
\\
\begin{acknowledgements} 
  We would like to thanks B. F. Svaiter for the useful discussions. This work was partially supported by Conselho Nacional de Desenvolvimento Cient\'{\i}fico e Tecnol\'{o}gico - CNPq, the grant - 301751/2019-6 (N.F.S). H.T.L.thanks Coordenação de Aperfeiçoamento de Pessoal de Nivel Superior (CAPES) for a MSc scholarship. G. O. H thanks Coordenação de Aperfeiçoamento de Pessoal de Nivel Superior (CAPES) for a Ph.D. scholarship.
\end{acknowledgements}

%

\end{document}